
\let\includefigures=\iftrue
%
%
\let\useblackboard=\iftrue
%
%
%
\input harvmac.tex
\let\expandedversion=\iffalse
\includefigures
\message{If you do not have epsf.tex (to include figures),}
\message{change the option at the top of the tex file.}
\input epsf
\def\fig#1#2{\topinsert\epsffile{#1}\noindent{#2}\endinsert}
\def\ebox#1#2{\topinsert\epsfbox{#1}\noindent{#2}\endinsert}
\else
\def\fig#1#2{\vskip .5in
\centerline{Figure 1}
\vskip .5in}
\def\ebox#1#2{\vskip .5in
\centerline{Figure 2}
\vskip .5in}
\fi
\def\Title#1#2{\rightline{#1}
\ifx\answ\bigans\nopagenumbers\pageno0\vskip1in%
\baselineskip 15pt plus 1pt minus 1pt
\else
\def\listrefs{\footatend\vskip 1in\immediate\closeout\rfile\writestoppt
\baselineskip=14pt\centerline{{\bf References}}\bigskip{\frenchspacing%
\parindent=20pt\escapechar=` \input
refs.tmp\vfill\eject}\nonfrenchspacing}
\pageno1\vskip.8in\fi \centerline{\titlefont #2}\vskip .5in}

\ifx\answ\bigans\def\tcbreak#1{}\else\def\tcbreak#1{\cr&{#1}}\fi
\useblackboard
\message{If you do not have msbm (blackboard bold) fonts,}
\message{change the option at the top of the tex file.}
\font\blackboard=msbm10 scaled \magstep1
\font\blackboards=msbm7
\font\blackboardss=msbm5
\newfam\black
\textfont\black=\blackboard
\scriptfont\black=\blackboards
\scriptscriptfont\black=\blackboardss
\def\Bbb#1{{\fam\black\relax#1}}
\else
\def\Bbb{\bf}
\fi
%
\def\l{\left}
\def\r{\right}
\def\comments#1{}

\def\BR{\Bbb{R}}
\def\BZ{\Bbb{Z}}
\def\p{\partial}

\def\half{{1\over 2}}
\def\Tr{{\rm Tr\ }}
\def\tr{{\rm tr\ }}
\def\Re{{\rm Re\hskip0.1em}}
\def\Im{{\rm Im\hskip0.1em}}

\def\vev#1{\langle{#1}\rangle}

\def\CA{{\cal A}}
\def\CC{{\cal C}}
\def\CF{{\cal F}}

\def\CM{{\cal M}}
\def\CN{{\cal N}}

\def\CL{{\cal L}}

\def\a{\alpha}
\def\bA{\bar A}
\def\floor#1{\lfloor {#1}\rfloor}

\Title{\vbox{\baselineskip12pt
\hfill{\vbox{
\hbox{hep-th/9503163 ~~~ RU-95-12\hfil}
}}}}
{\vbox{\centerline{Dynamics of $SU(N)$}
\medskip\centerline{Supersymmetric Gauge Theory}}}
\centerline{Michael R. Douglas}
\smallskip
\centerline{\it and}
\smallskip
\centerline{Stephen H. Shenker}
\smallskip
\centerline{Dept. of Physics and Astronomy}
\centerline{Rutgers University }
\centerline{Piscataway, NJ 08855-0849}
\centerline{\tt mrd@physics.rutgers.edu}
\centerline{\tt shenker@physics.rutgers.edu}
\bigskip
\bigskip
\noindent
We study the physics of the Seiberg-Witten and
Argyres-Faraggi-Klemm-Lerche-Theisen-Yankielowicz solutions of $D=4$, $\CN=2$
and $\CN=1$ $SU(N)$
supersymmetric gauge theory.
The $\CN=1$ theory is confining and its effective Lagrangian is a spontaneously
broken $U(1)^{N-1}$ abelian gauge theory.
We identify some features of its physics which see this internal
structure, including a spectrum of different string tensions.
We discuss the limit $N\rightarrow\infty$, identify a scaling regime in
which instanton and monopole effects survive, and
give exact results for the crossover from weak to
strong coupling along a scaling trajectory.  We find a large hierarchy
of mass scales in the scaling regime, including very light
$W$ bosons, and the absence of weak coupling.  The light $W$'s leave
a novel imprint on the effective dual magnetic theory.
The effective Lagrangian appears to be inadequate to understand the
conventional large $N$ limit of the confining $\CN=1$ theory.

\Date{March 1995}
\nref\sei{N. Seiberg, Nucl. Phys. B435 (1995)129,
hep-th/9411149 and
references therein.  For a brief review, see
``The Power of Holomorphy,'' N. Seiberg, hep-th/9408013.}
\nref\sw{N.~Seiberg and E.~Witten, Nucl.Phys. B426 (1994) 19.
The generalization to theories with  matter hypermultiplets is discussed in
N.~Seiberg and E.~Witten, hep-th/9408099.}
\nref\AF{P.~C.~Argyres and A.~E.~Faraggi, hep-th/9411057.}
\nref\KLTY{A.~Klemm, W.~Lerche, S.~Yankielowicz and S.~Theisen,
hep-th/9411048 and hep-th/9412158.}
\nref\oldsei{N.~Seiberg, Phys.~Lett. 206B (1988) 75.}
\nref\tH{G.~'t Hooft, Nucl.~Phys. B190 [FS3] (1981) 455.}
\nref\vafaetal{S. Cecotti, P. Fendley, K. Intriligator, and C. Vafa,
Nucl. Phys. B386 (1992) 405; S. Cecotti and C. Vafa, Comm. Math. Phys.
158 (1993) 569.}
\def\numdefC{(2.6)}
\def\Bdiag{(A.1)}
\def\qsin{(A.5)}
\newsec{Introduction}

Over the last year and a half, revolutionary progress in understanding
the dynamics of four-dimensional supersymmetric gauge theories has been
made by Seiberg and collaborators~\refs{\sei}.  One
spectacular result
is the exact low-energy
effective Lagrangian for $\CN=2$ supersymmetric $SU(2)$
gauge theory obtained by Seiberg and Witten~\sw\ .
The $\CN=2$ vector multiplet contains an adjoint scalar whose
non-zero vacuum expectation value (vev) breaks $SU(2)$ to $U(1)$.  For
large vev compared to the scale $\Lambda$ set by the gauge coupling, one
can write an effective Lagrangian in terms of a $U(1)$ gauge multiplet.
For small vev, in the past we would have expected $SU(2)$ color
confinement and a very different (and inaccessible) description.

Surprisingly, it turned out that the effective theory is a $U(1)$ gauge theory
for any expectation value, and the naive unbroken $SU(2)$ regime is not present
at all.  Instead, two singularities of the effective action appear in the
strong coupling regime.
At one of these, the monopole visible in the semiclassical treatment becomes
arbitrarily light.
The effective Lagrangian is again a $U(1)$ gauge theory but now written in
terms of a vector multiplet containing the dual (magnetic) gauge field with the
standard local coupling to the monopole, as well as a hypermultiplet describing
the monopoles.
The other singularity is isomorphic to this,
with the role of the monopole taken by a charge $(1,1)$ dyon.

Thus the $\CN=2$ theory does not confine.
Surprising as this result may be, it does not drastically contradict previous
expectations, mostly because of the presence of the massless scalar.
Theories which do not have such a scalar and are widely believed to be similar
to pure (bosonic) gauge theory are $\CN=1$ SYM gauge theory as well as
supersymmetric QCD.  These theories should confine, and Seiberg and Witten
showed that this could be explained in $\CN=1$ theories obtained by adding an
$\CN=2$ breaking mass term: it is the result of monopole condensation.

These results have been extended by Argyres and Faraggi \AF\ and by
Klemm, Lerche, Theisen and Yankielowicz~\KLTY\ to pure $SU(N)$ gauge
theory.\footnote*{ It should be noted that the precise dependence of the
effective action on the moduli proposed in equations (4) and (14) of the
second paper in \KLTY\ is ambiguous for general $N$.  We follow the
unambiguous results
in equations (6) and (9) of \AF\ .} The $\CN=2$ theory now has $N-1$ moduli
(say, the eigenvalues of the adjoint scalar vev $\vev{\phi}$) and in the
semiclassical regime the gauge symmetry is broken to $U(1)^{N-1}$.  Each
factor contains a monopole solution and by varying the moduli one can
drive any of them massless.  There are $N$ points in moduli space
at which $N-1$ monopoles becomes massless, and these points become the
ground states of the associated $\CN=1$ theory.

In this paper we continue the discussion of the physics of these
theories begun in \refs{\sw,\AF,\KLTY}.
Perhaps the most striking result is the following.
In the weakly coupled regime, the theory is a $U(1)^{N-1}$ gauge theory with a
discrete gauge symmetry $S_N$ permuting the $U(1)$ factors.
This discrete gauge symmetry is spontaneously broken by the Higgs vev -- for
example, at a generic point in moduli space, every charged multiplet has a
distinct mass.
It turns out that this is true everywhere in moduli space,
even at the vacua of the $\CN=1$ theory.  This leads to a non-trivial spectrum
of light massive particles in the $\CN=1$ theory and rather surprisingly, a
spectrum of distinct string tensions in the different $U(1)$ factors.
Since the theory contains particles with charges in any pair of $U(1)$ factors,
the limiting (infinite distance) string tension is the lowest of these, but
mesons and baryons bound with the higher string tensions will exist as
sharp resonances for sufficiently small $\CN=2$ breaking.

We also consider the large $N$ limit of the theory and compare with
expectations from previous work.  Let us first say a few words about
possible large $N$ limits.  The bare coupling constant is always taken
to zero as $g_0^2\sim 1/N$ to get a theory with a planar diagram
expansion.  Now we will discuss only the leading (two derivative)
effective Lagrangian, and in $\CN=2$ SYM this receives no perturbative
contributions beyond one loop.  Since instantons are suppressed as
$e^{-N/g_0^2}$, it may at first sound like there is little to do.
However, this statement turns out to be naive.

In the $\CN=2$ theory, there are two natural regimes to consider.  To
give all charged bosons masses of at least $O(N^0)$, the difference between any
pair of eigenvalues of $\phi$ must be at least $O(N^0)$, and the typical
difference will be $O(N)$.\footnote*{We assume for simplicity that all
eigenvalues are real.  Otherwise there are configurations where the
typical difference will be $O(\sqrt{N})$.} The density of states will be
$O(N)$.  We refer to this as the {\it naive semiclassical regime} as
monopole masses are $O(N)$, and instanton corrections do not survive in
this limit.  The cause of this is {\it not} only the  $e^{-N/g^2}$
suppression.  Rather it is a non-trivial consequence
of the need to introduce $N$ distinct mass scales.

To make instanton corrections survive, one must take the spacing between
eigenvalues to be $O(1/N)$.  In this {\it scaling regime}, monopoles are light.
The vacua of the $\CN=1$ theory are of this form, and one can smoothly
interpolate between them and a limit in which the spacings are $\lambda/N$ with
$\lambda$ becoming large, in which the semiclassical treatment is valid.
In this sense, there is no large $N$ transition in the theory.

Since the effective theory is a spontaneously broken abelian gauge theory,
the standard expectations of the large $N$ limit -- a finite mass gap,
$O(1)$ degeneracies in the particle spectrum, and an effective $\phi^3$
coupling of order $1/N$ -- are not at all obvious.  We find that at least the
last of these is violated near the massless monopole point.

The traditional definition of the large $N$ limit takes $N\rightarrow\infty$
before any other limits and in particular before the infinite volume limit.
Our low energy effective Lagrangian should be accurate for processes involving
momenta $|p|<<\Lambda_{eff}$, where the scale is set both by the masses of
particles we have integrated out and by the couplings of higher derivative
terms we have dropped.
Electrically charged particles must be integrated out, and it will turn out
that the lightest of these has mass $\sim \Lambda/N^2$, which will make the
interpretation of the limit subtle.

\newsec{$SU(N)$ supersymmetric gauge theory}

The $\CN=2$-supersymmetric bare Lagrangian is
(in $\CN=1$ superfield notation)
\eqn\slag{\CL = \Im {N\tau_0\over 4\pi}\left[~
\int d^4\theta A^a\bA^a + \int d^2\theta W_\a^a W_\a^a~\right].}
$\tau_0=4\pi i/g_0^2 + \theta/2\pi$ is the bare gauge coupling.
There are two reasons for the explicit $N$ dependence.
Of course, it is the appropriate one to
weigh Feynman diagrams of Euler character $\chi$ as $N^\chi$,
and thus we will take the large $N$ limit with $\tau_0$ fixed.
It is convenient even at finite $N$, since it cancels the explicit $N$ in the
one-loop beta function, and thus the dynamical scale
$\Lambda\sim \exp-\Im\tau_0$ (at which the running coupling attains a
prescribed $O(1)$ value)
will have no $N$ dependence.

The $\CN=2$ theory has an $N-1$ complex dimensional moduli space $\CM$
of vacua.
In the classical theory these are
parameterized by the invariant expectation values $\Tr\phi^n$ constructed from
the scalar component of $A$.
$\phi$ must satisfy the $D$-flatness condition $[\phi,\phi^+]=0$ and thus we
can diagonalize it, and use as coordinates for $\CM$ the eigenvalues $\phi_m$
with permutations identified.
At a generic point $\phi$ breaks the gauge symmetry to $U(1)^{N-1}$ and
a low-energy effective Lagrangian can be written in terms of multiplets
$(A_i,W_i)$.  We will use a `$U(1)^N$' notation in which $1\le i\le N$ and
$\sum_i A_i = 0$.  We denote the scalar component of $A_i$ by $a_i$.

The $\CN=2$ effective Lagrangian is determined by an analytic prepotential
$\CF$ and takes the form
\eqn\effL{\CL_{eff}=\Im{{1}\over{4 \pi}} \left[
\int d^4\theta\ \p_i \CF(A)\bA^i + \half\int d^2\theta\ \p_i\p_j\CF(A) W^i W^j
\right].}
In the classical theory,
$\CF_{cl}(A) = {N\tau_0 \over 2}\sum_i (A_i-\sum A_j/N)^2$.
For large $\phi_m$ the gauge coupling is weak at the scale of symmetry breaking
and a good approximation to $\CF$ would be obtained by adding the one-loop
quantum correction:
\eqn\Foneloop{\CF_1 = {i\over 4\pi}
\sum_{i<j} (A_i-A_j)^2 \log {(A_i-A_j)^2 \over e^3\Lambda^2}.}
One can renormalize and define $\Lambda$ to absorb $\CF_{cl}$ into this
expression.\footnote*{
We introduced the extra `$e^3$' compared with \AF\ to be consistent with
an extra $\half$ in the curve \numdefC\ below.
It will turn out that this simplifies the formulas at the massless monopole
point.}
In fact, $\CN=2$ supersymmetry forbids further perturbative
corrections~\oldsei\ . %
This is not to say that perturbation theory is trivial but that it will only
produce higher derivative terms in the effective Lagrangian.

The reduced (or BPS saturated) multiplets and their mass formula play a central
role in the story.
The $U(1)^{N-1}$ theory has a lattice of allowed electric and magnetic charges,
which we write $q_i$ and $h_i$, again with $\sum_i q_i=\sum_i h_i=0$.
The charges of vector bosons are the vectors
$q_v=(0,\ldots, 0,+1,0, \ldots, 0,-1,0,\ldots)$.
The fundamental representation (we use the convenient name `quark') are
$q_q=(0,\ldots, 0,1,0\ldots 0)-(1/N,\ldots, 1/N)$.
The theory contains 't Hooft-Polyakov monopoles as well, whose magnetic charges
 must satisfy the DSZ condition
$q^{(1)}\cdot h^{(2)}-q^{(2)}\cdot h^{(1)}\in\BZ$.
If we order the eigenvalues $\phi_i>\phi_{i+1}$, we expect the stable monopoles
to be those with charges in successive factors:
$h=(0,\ldots, 0,+1,-1,0,\ldots)$, so we introduce the basis vectors
$h_m^i=\delta_m^i-\delta_{m+1}^i$.
The mass of a reduced multiplet is determined by its $\CN=2$ central charge,
which is determined by its electric and magnetic charges to be
\eqn\cent{M = \sqrt{2}|Z| =  \sqrt{2}|a\cdot q + a_D\cdot h|}
with $a_{Di}=\p\CF/\p a_i$, a result motivated by duality as we discuss below.

The $a_D$ associated with individual monopoles are
$a_{Dm}=h_m^i a_{Di}$, whose inverse is
$a_{Di} = a_{D,i=1}-\sum_{m<i} a_{Dm}$.
We will always use the indices $ij$ versus $mn$ to distinguish the
two bases.
Finally, we introduce $a_n=\sum_{i\le n} a_i$ with inverse
$a_i=q_i^n a_n=a_{n=i}-a_{n=i-1}$.
These are canonically conjugate to the $a_{Dm}$
in the sense that $a_n=-\p\CF_D/\p a_{Dn}$.

We first discuss the naive semiclassical regime and its large $N$ limit.
To break $SU(N)$ to $U(1)^{N-1}$ and give every charged multiplet a mass
of order $N^0$ or greater, we need to choose $\phi$ so that the minimum
difference between eigenvalues is $O(N^0)$.  A representative choice is
$\phi_{ij}=v (i-(N+1)/2)\delta_{ij}$, where $v$ is the characteristic
scale of the vev.  The classical prediction for the mass of the
multiplet $A_{ij}$ is then $v|i-j|$ and we have a linearly rising
spectrum with multiplicity $O(N)$ at each mass level.  Monopoles are
expected to have masses $4\pi N v/g_0^2$ and do, although it may be
interesting to note that the formula $a_{Di}=\p(\CF_{cl}+\CF_1)/\p a_i$
produces unusual corrections to this, e.g. with $\log N$ dependence.

We now turn to the exact solution \refs{\sw,\AF,\KLTY} .
$\CF(A)$ is determined in the quantum theory by combining analyticity with a
physical ansatz for the number and types of its singularities, at points where
BPS saturated states become massless.
What makes it possible to study the strong coupling regime and combine the
information from different regimes is
the existence of exact duality transformations on the effective Lagrangian,
and the generalization of the Witten effect:
encircling any singularity in moduli space produces a non-trivial
$Sp(2N-2;\BZ)$ transformation on the electric and magnetic charges of all
states.

$\CF(A)$ is not a single-valued function of $A$ (as is already clear from
\Foneloop) and thus we must distinguish the coordinates on $\CM$ (for which we
retain the name $\phi_i$) from the scalar components $(a_{Di},a_i)$ of the
fields in the effective Lagrangian.
It turns out that $\CF(A)$ is most simply expressed in terms of an auxiliary
Riemann surface $\CC$ which varies on $\CM$, defined by the curve
\eqn\defC{\eqalign{
y^2 &= P(x)^2 - \Lambda^{2N}\cr
P(x) &\equiv \half\det~(x-\vev{\phi}) = \half\prod_i (x-\phi_i).}}
(We generally set $\Lambda=1$ in the following).
The $(a_{Di},a_j)$ are then integrals of the meromorphic
form $\lambda=(1/2\pi i)(x/y) dP(x)$ over a basis of one-cycles with the
intersection form $h_i\cdot q_j$.

To define the one-cycles, order the branch points $x_i$.
Let $\gamma_i$ for $1\le i\le N$
encircle the branch points $x_{2i-1}$ and $x_{2i}$,
and $\alpha_m$ for $1\le m< N$ encircle the cut running from
$x_{2m}$ to $x_{2m+1}$.  The $\gamma_i$ are not independent but satisfy
$\sum_i\gamma_i=0$, while the $\alpha_m$ are independent.
Their intersection matrix is
$\vev{\alpha_m,\gamma_j}=\delta_{m,j}-\delta_{m,j+1}$
and thus we can associate the quarks with the cycles $\gamma_i$
and the monopoles with $\alpha_m$.
We also define a set of cycles conjugate to the $\alpha_m$ as
\eqn\cycledef{\beta_n=\sum_{i\le n}\gamma_i.}
We then write
\eqn\periods{a_{Dm} = \oint_{\alpha_m}\lambda \qquad\qquad
a_{n} = \oint_{\beta_n}\lambda.}

\fig{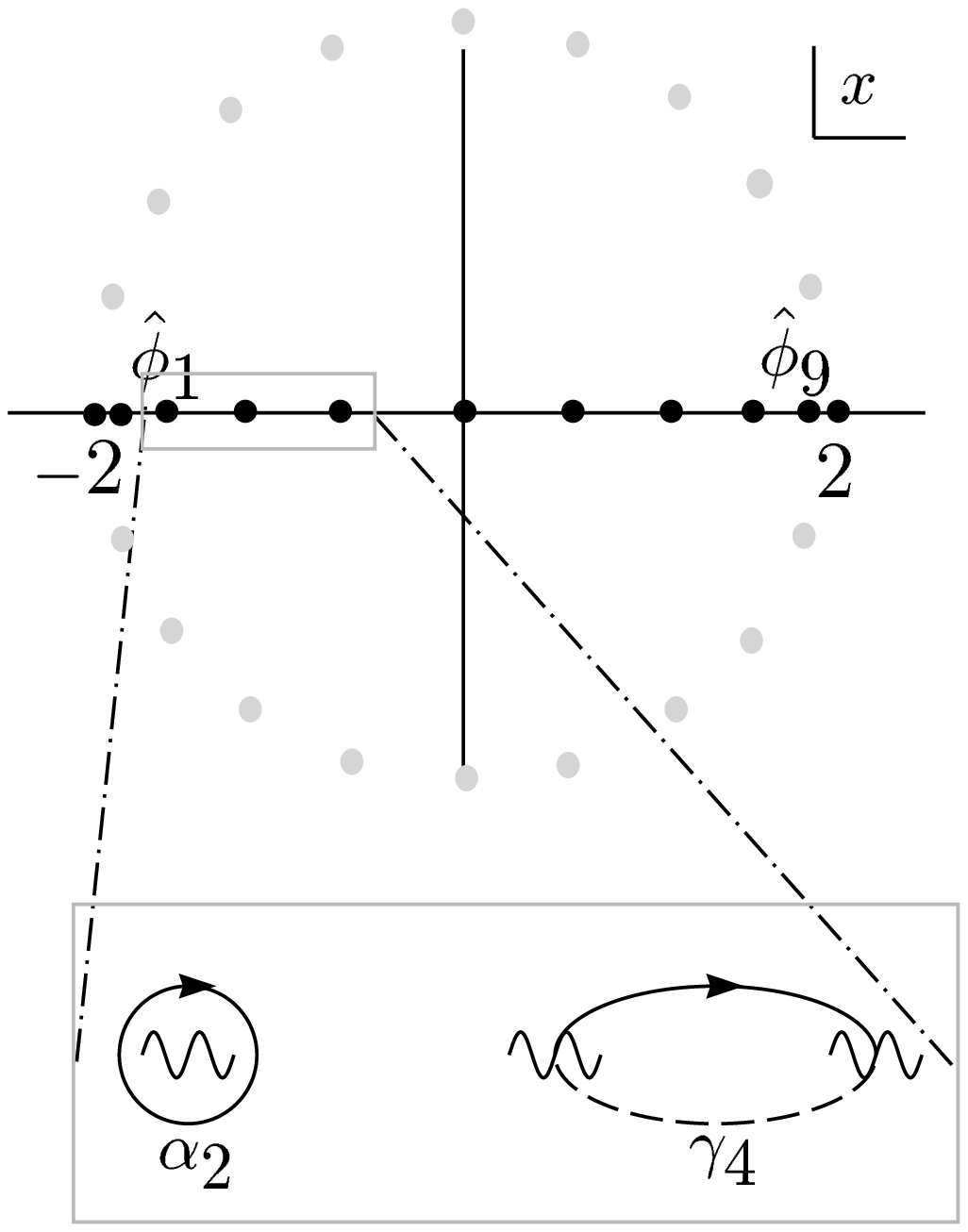}{}

The strong coupling regime is controlled by points where monopoles and dyons
become massless: $\vec h\cdot \vec a_D+\vec q\cdot\vec a=0$.
The vacua of the $\CN=1$ theory will come from points in moduli space at which
monopoles coupling to each $U(1)$ become massless.
There are $N$ such points, with a simultaneous degeneration of all the
$\alpha$-cycles of the quantum curve.  In the realization $y^2=P(x)^2-1$ this
will happen when the $N$ cuts are lined up, each with one branch point
coinciding with the next.
In other words, we require $P(x)^2-1$ to have $N-1$ double zeros and two single
zeros.  This condition can be satisfied using Chebyshev polynomials:
\eqn\cheb{\eqalign{
&P(x) = \half\det~(x-\vev{\phi}) = T_N\l({x\over 2}\r) =
\cos\l(N \arccos{x\over 2}\r)\cr
&P(x)^2-1 = \l({x^2\over 4}-1\r)~ U_{N-1}\l({x\over 2}\r)^2.}}
We obtain $N-1$ more solutions by complex rotations
$x\rightarrow e^{i\pi r/N} x$.  Since each is associated with a ground state of
the $\CN=1$ theory, whose Witten index is $N$, we do not expect other solutions
to exist.\footnote*{
The full story is  subtle as there are partial degenerations of the curve
at which $N-1$ periods of $\lambda$ vanish.
These are limits in which more than two branch points coalesce, and
an example is $P(x)=\half x^N-\Lambda^N$.
A coalescence of $k$ branch points will cause the periods of $\lambda$ on
cycles surrounding any two branch points to vanish simultaneously.  One can
choose a basis of $k-1$ such cycles, and they have a non-zero intersection
form, meaning that the associated massless particles would not be mutually
local. One can show that at such points at least one $U(1)$ will not couple to
any of the massless particles and so will not confine when an
$\CN=2$-breaking mass perturbation forces the massless particles to condense.
Thus these are not candidates for $\CN=1$ vacua,
where full confinement is expected.
We thank P. Argyres, W. Lerche and L. Randall for discussions on this point.}

The eigenvalues of $\phi$ are non-degenerate, $\phi_n = 2 \cos \pi(n-\half)/N$.
They have spacing $O(1/N)$ in the center of the ``band'' and $O(1/N^2)$
at the edges.
The double branch points of the curve are $\hat\phi_n = 2 \cos \pi n/N$
for $1\le n<N$ and the single branch points are at $\pm 2$.

Let us compute the periods of the maximally degenerate curve $\CC_0$.
Since the cuts are all on the real axis, the $\alpha$ periods will be imaginary
while the $\gamma$ periods will be real.
Changing variables from $x=2\cos\theta$ to $\theta$, we have
\eqn\perstuff{\eqalign{
P(x)&= \cos N\theta\cr
y&= i\sin N\theta\cr
\lambda &= {1\over 2\pi i}{x\over y}{\p P(x)\over\p x} dx\cr
&= {N\over\pi} \cos\theta d\theta\cr
{\p \lambda \over\p\phi_i}\bigg|_{\CC_0}&=
-{1\over 2\pi i}{1\over y}{\p P(x)\over\p\phi_i} dx ~+d(\ldots)\cr
&= {1\over\pi}{1\over x-\phi_i}\cot N\theta \sin\theta~d\theta.}}

The $a_{Dm}$ are integrals around the $\alpha$ cycles.
 These degenerate to integrals around the $\hat\theta_m$.
Since $\lambda$ is non-singular, all $a_{Dm}=0$ for $\CC_0$.
When we vary the curve, we could get two types of contributions.
First, the zeroes of $P^2-1$ split.  This will be important below
but here we simply inflate the contour to enclose the new branch points.
Second, the derivatives of $\lambda$ have poles:
\eqn\perstufftwo{\eqalign{
B_{mi} \equiv -i{\p a_{Dm}\over\p\phi_i} &= -{i\over
\pi}\oint_{\alpha_m}d\theta~
{1\over x-\phi_i}\cot N\theta\sin\theta\cr
&= {2\over \phi_i-\hat\phi_m}~
{\cos N\theta\sin\theta\over{\p\over\p\theta}\sin N\theta}
\bigg|_{\theta=\hat\theta_m}\cr
&= {1\over N}{\sin\hat\theta_m\over \cos\theta_i-\cos\hat\theta_m}~
}}
The matrix $B_{mi}$ is simple in the following basis (appendix A):
$$\sum_{m=1}^{N-1} B_{mi} \sin {\pi km\over N} =
 \cos {\pi k(i-\half)\over N}.\eqno(A.1)$$

The $a_{m}$ are integrals around the $\beta$ cycles.
We find
\eqn\perstuffthree{\eqalign{
a_m &= 2\int_{0}^{\hat\theta_m}\lambda\cr
&= {2N\over\pi}\int_{0}^{\pi m/N}d(\sin\theta) \cr
&= {2N\over\pi}\sin {\pi m\over N}.}}

These determine the masses of BPS saturated electrically charged states.
Quark masses would be given by the $\gamma$ periods, which are differences
of $\beta$ periods:
$m_j={2\sqrt{2}N\over\pi}
(\sin {\pi j\over N}-\sin {\pi (j-1)\over N})$.
For odd $N$, $m_{(N+1)/2}$ vanishes, but since there are no quarks in the
theory at hand, this is not a difficulty.
The `W bosons' which are present have masses $m_{ij}=m_i-m_j$,
which are all non-zero at finite $N$.

The first comment to make is that these masses are different for particles with
charges in different $U(1)$ factors.  As we commented in the introduction, this
is generically true at weak coupling: an explicit choice of Higgs vev will
break the discrete gauge symmetry relating the factors.  What may be surprising
is that this persists for small Higgs vev and at the vacuum relevant for the
related $\CN=1$ gauge theory.  There is simply no symmetry under permuting the
periods of the curve $\CC_0$.

Our second comment is that the mass of the
lightest $W$ boson goes to zero in the large $N$ limit as $m_{12}=\Lambda
\pi^2/N^2$.
This is important because it determines the energy scale at which our effective
Lagrangian breaks down, and we discuss this point below.

The derivatives of the $a_{m}$ are
\eqn\perstufffour{\eqalign{
A_{mi} = {\p a_{m}\over\p\phi_i} &= {1\over 2\pi}\oint_{\beta_m}d\theta~
{\sin\theta\over \cos\theta-\cos\theta_i}\cot N\theta\cr
&= {1\over \pi}\int_{0}^{\pi m/N}d\theta~
{\sin\theta\over \cos\theta-\cos\theta_i}\cot N\theta\cr
}}
This is log divergent, as it should be on physical grounds.
The dual gauge coupling in each $U(1)$ factor has a positive (IR free) beta
function produced by a one-loop diagram involving its light monopole.
Away from the massless monopole point,
it will be cut off at the mass of the monopole to
produce the same result in each factor as in \AF:\footnote*{
The $i/\pi$ of \sw\ is in different charge conventions.}
$4\pi/e_D^2\sim\tau_{mn}^D \sim -(i/2\pi)\delta_{mn}\log a_{Dm}$.

To get a finite result one must perturb the curve slightly to
give the monopoles small masses.  This again splits the double zeroes
and modifies $\lambda$.  For calculating the coefficient of the logarithm, the
modification of $\lambda$ does not matter, and the result is given by
integrating $(\p\lambda/\p\phi_i)_{\CC_0}$ up to the branch point, which we
parametrize as
$\theta\equiv\hat\theta_m-\epsilon_m/N$.
The integral then simply gives an endpoint divergence,
\eqn\perstufffoura{\eqalign{
A_{mi} &\sim -{1\over \pi N}
{\sin\theta\over \cos\theta_i-\cos\theta}\log\sin N\theta
\bigg|_{\theta=0}^{\theta=\pi m/N-\epsilon_{m}/N}\cr
&\sim -{1\over \pi}\log\epsilon_m~ B_{mi}.}}

The $\epsilon_m$ are computed by varying the equation $y^2=P(x)^2-1$.
Since we are splitting double zeroes we will find
$\epsilon\sim(\delta\phi)^{1/2}$.
Using \perstuff\ we have
\eqn\splits{\eqalign{
0 &= (\delta y)^2 + 2P\sum_i{\p P\over\p\phi_i}\delta\phi_i\cr
&= - \sin^2 N\delta\theta - 2 \cos^2 N\theta
 \sum_i{\p P\over\p\phi_i}\delta\phi_i\cr
\epsilon_m^2 &= -2 \sum_i {1\over\hat\phi_m-\phi_i}\delta\phi_i\cr
&= {N\over\sin\hat\theta_m} \sum_i B_{mi} \delta\phi_i\cr
&= {N\over\sin\hat\theta_m} \delta a_{Dm}}}
using \perstufftwo.

Thus the period matrix diverges as
\eqn\perstufffourb{\eqalign{
\tau^D_{mn}&={\p a_m\over\p a_{Dn}}
=-i\sum_i A_{mi}B^{-1}_{ni}\cr
&\sim -{i\over 2\pi}\delta_{mn}~\log {a_{Dm}\over \Lambda_m}}}
with $\Lambda_m\equiv \Lambda\sin\hat\theta_m/N$.\footnote*{
The preceding calculation of $A_{mi}$
left out constant terms coming from the bulk of the integral,
which could change
$\Lambda_m$.  In section 5 we will do this more carefully and show that
important constants are present, but are compatible with the definition of
$\Lambda_m$ we give here.}
This checks with the expectations for the beta function, and
confirms the existence of one monopole in each factor
(there is no overall $N$).

At this order the different $U(1)$ factors are completely decoupled.
Note that the decoupled factors are different at weak and at strong coupling
($\tau$ is diagonal in a different basis).
Physically, the two limits are controlled by different light degrees of
freedom.
The $m$'th monopole beta function turns on at the scale $\Lambda_m$, and
at sufficiently low energy, their contribution $\delta_{mn}/e_D^2$
will dominate any structure in $\tau$ from higher
energies, and decouple the factors.

In general, an extended object of size $L$ does
not contribute to the beta function at energies $E>1/L$, so we can
interpret the scale $\Lambda_m$ as an indication of the size of
the monopole.  Semiclassically the monopole size would have been
$L\sim 1/m_W$ and even though here gauge couplings are $O(1)$,
we still have $\Lambda_m= m_{m,m+1}/2\sqrt{2}\pi$
with $m_{m,m+1}$ to good accuracy the
lightest $W$ mass coupling to that factor.

The effective Lagrangian around the maximal degeneration is
\eqn\maxdeg{\eqalign{
\CL_{eff} = \sum_m \bigg[~ \Im {i\over e_{Dm}^2}&\left(~
\int d^4\theta A_D^m\bA_D^m + \int d^2\theta (W_D^m)^2~\right)\cr
&+ \Im \int d^4\theta M_m^+ e^{V_{Dm}} M_m
+ \tilde M_m^+ e^{-V_{Dm}} \tilde M_m\cr
&\qquad+ \Re \sqrt{2}\int d^2\theta A_{Dm} M_m \tilde M_m ~\bigg].}}
The $U(1)$ factors will couple at higher orders in an
expansion of the kinetic term, but we expect these interactions to
be suppressed by powers
of $p/\Lambda_m$.  One physical source of this coupling is loops of
massive $W$ particles charged under more than one $U(1)$, and such
interactions will be suppressed by the mass of the $W$s.

An interesting question for contact with the standard large $N$ limit (to
be discussed below) is whether the interactions are suppressed in the standard
way, as $\phi^{2+k}/N^k$.  It is obviously not so for the superpotential terms
in \maxdeg, which are completely determined by $\CN=2$.  Nor is there an
obvious reason for it to be true of the couplings between $U(1)$ factors.

A $\CN=1$ theory can be obtained by adding the perturbation $W=Nm\Tr A^2$
to the bare superpotential.
Following \sw\ we interpret this in the effective Lagrangian as the
perturbation $W=Nm\sum_i \phi_i^2$, the observable (single-valued on moduli
space) equivalent to $Nm\Tr A^2$ in the semiclassical limit, and thus
the exact superpotential is
\eqn\exactsup{W = \sqrt{2}\sum_m A_{Dm}M_m\tilde M_m + Nm\sum_i
\phi_i^2.}

To calculate its effect we need to change variables from $\phi_i$ to $a_{Dm}$.
We find that the vacua $W'=0$ satisfy
\eqn\noneeff{\eqalign{
\sqrt{2}\vev{M\tilde M}_n = 2Nm\sum_i (B^{-1})_{ni} \phi_i\cr
\sum_n B_{ni} \vev{M\tilde M}_n = \sqrt{2}Nm\phi_i}}
and $a_{Dn}=0$.
This means of course that the kinetic term cannot literally be
given by \perstufffourb.
Physically, the monopole loop integrals are now cut off by masses
produced by the $\CN=1$ part of the superpotential.  We use the
prescription of taking $a_{Dn}=\vev{M\tilde M}_n^{1/2}$ in the $\CN=2$
kinetic term to account for this.

The equation \noneeff\ has the solution
\eqn\nonesol{\vev{M\tilde M}_n = 2\sqrt{2}Nm \sin {\pi n\over N}}
and all of the $U(1)$ gauge symmetries are spontaneously broken.
We did not find an explicit discussion of the particular $\CN=2$ abelian Higgs
theory which appears here in the literature, but it is easy to work
out\footnote*{One can simplify one's life by observing that the linear part of
the $\CN=2$ breaking term in $W$ is a pure $F$ term, which using $\CN=2$ can be
rotated into a $D$ term.  This turns the vev into
$\vev{M^2}=2\sqrt{2}Nm \sin {\pi n\over N}$ and $\vev{\tilde M}=0$ and the
string solution becomes that of the usual $\CN=1$ abelian Higgs model.}
that the particles in a given $U(1)$ factor all have the same mass
\eqn\massfor{m_n^2= 2 e_{Dn}^2 \vev{M\tilde M}_n}
and the string tension in the factor is
\eqn\stringten{\kappa_n = 2 \pi \vev{M\tilde M}_n.}
It is different in each factor,
and roughly proportional
to the lightest $W$ mass in the factor, $\kappa_n\propto m N^2 m_{n,n+1}$.
The gap is non-vanishing in the large $N$ limit.
Because the scalar and vector retain the same mass after $\CN=2$ breaking,
there is no long range potential between two strings (they are `neutrally
stable').

Using our prescription to determine the kinetic term,
\eqn\kinstu{\eqalign{
{4 \pi\over e_{Dn}^2} &\sim -{1\over 4\pi}\log( {m\Lambda\over\Lambda_n^2}
N \sin {\pi n\over N})\cr
&= -{1\over 4\pi}\log ({mN^3\over \Lambda\sin {\pi n\over N}})}}
and there is a weak (logarithmic) variation between the couplings in the
different $U(1)$ factors.

\newsec{Physics at finite $N$.}

The effective Lagrangian \maxdeg\ with the superpotential \exactsup\ %
provides an explicit realization of the abelian monopole condensation
model of confinement in an $SU(N)$ gauge theory.
Such models of confinement were much discussed previously, as in \tH,
on a qualitative level.
However, what in retrospect seems an obvious question was not discussed
to our knowledge: namely,
how can an analysis in which the theory looks like a broken $U(1)^{N-1}$ gauge
theory avoid having $N-1$ distinct types of flux tubes and a corresponding
multiplicity in the spectrum?

In fact we see from \stringten\ that
the $U(1)$ factors have differing string tensions.
(There is a $\BZ_2$ symmetry $a_i\rightarrow -a_{N+1-i}$,
$\vev{M\tilde M}_n\rightarrow\vev{M\tilde M}_{N-n}$ which reduces
the number of distinct string tensions to $\floor{N-1\over 2}$.)
Although the individual factors cannot be studied in isolation,
since the coupling between them is due to the finite mass $W$ bosons,
one can see physical effects of the `heavier' factors.

The simplest example is the expectation value of a Wilson loop.
At sufficiently long distances it is energetically favorable for the heavier
string tensions to be screened by $W$ bosons.
The parameter which controls this is the ratio of string tensions to $W$
masses $\kappa/m_W^2\sim N^2 m/\Lambda$, and for any $m/\Lambda<<1$
the almost-$\CN=2$ analysis should be valid.
Thus in the area law regime $\kappa L^2>>1$, the Wilson loop will show a
crossover from an intermediate distance $L<m_W/\kappa\sim 1/N^2 m$ behavior,
the sum of
terms $\exp -\kappa_n L^2$ with distinct string tensions, to a long distance
behavior governed by the lowest string tension.

The distinct string tensions are already visible for $SU(3)$.
Since the quark charges expressed in the monopole basis are
$(1\ 0)$, $(-1\ 1)$ and $(0\ -1)$, the intermediate range fundamental
Wilson loop will go as $2 \exp -\kappa_1 L^2 + \exp -2\kappa_1 L^2$.

The underlying $U(1)^{N-1}$ symmetry of the effective
theory has other physical consequences as well.
It is clearly visible in the
light spectrum, the `glueballs' described by the fields of our
effective Lagrangian.  These are weakly coupled because $e_D$ is small (for
$m<<\Lambda$), and the different
$U(1)$ factors are derivatively coupled (again, controlled by
the $W$ masses).

Heavier confined states will vary in each factor as well.
Let us consider a $q\bar q$ meson state.
(We assume for illustrative purposes that the theory with quarks is
qualitatively similar, which will be true for heavy enough quarks.
It is even true for light quarks in some cases, for example
$SU(2)$ with one flavor.
Of course one could talk about states containing gluinos
in the present theory.)
For each single $q\bar q$ state found in a conventional analysis (for example
the strong coupling expansion), it appears that we would find $N$ states,
distinguished by the charge of the quark,
and bound with different string tensions.

The $SU(N)$ gauge theory broken to $U(1)^{N-1}$ theory has an $S_N$
discrete gauge symmetry, and one might have thought that this
symmetry would somehow remove the multiplicity.
However,
we found that in the model under study, the dynamics picked a vacuum
$\vev{\phi_i}$ which spontaneously breaks the symmetry.
This affects the expectation values $a_i$ and the superpotential \exactsup,
and thus the string tensions and the physically observable `glueball' masses
\massfor\ all break the symmetry.
The masses of the $N$ meson states contain contributions
{}from both the string tensions and the quark
masses $m_q+a_i$, and (except possibly at special values
of $m$, $m_q$ and $N$), there are no degeneracies between them.
This leaves no doubt that they are distinct physical states.

The $N$ states
are not distinguished by any conserved quantum numbers and thus the
heavier states are unstable to decay into the lightest.
This is mediated by the derivative couplings between the $U(1)$ factors
and thus a decay amplitude will be
$\CA\sim (\Delta E)^2/m_W^2$.  From the string tensions
we can estimate $\Delta E\sim\sqrt{m\Lambda}$ and thus
the decay rate $\CA\sim m/\Lambda$ is controlled by the same
arbitrarily small parameter which allowed us to see the distinct
string tensions.

The structure of $N$-fold split multiplets is a rather unexpected
difference between the physics of `almost-$\CN=2$' supersymmetric gauge
theory and non-supersymmetric gauge theory.  Although definitive results
for pure $\CN=1$ gauge theory (or large $m$ in the present theory) do
not exist, we know of no evidence that this structure
would persist, and believe
it does not.  Rather, we believe it is associated with the special
features of almost-$\CN=2$ supersymmetry, in particular the light
scalar.

One indication of this is that the natural gauge-invariant
operators which create
these states distinguish the $U(1)$ factors by using the scalar vev.
They are the chiral superfields
\eqn\defok{O_k =\tilde Q P_k(\phi) Q.}
The polynomials $P_k(\phi)=c_k\prod_{i\ne k}(\phi-\phi_i)=2c_k
T_N(\phi/2)/(\phi-\phi_k)$
satisfy $P_k(\phi_i)=\delta_{ik}$ and pick out a single quark charge.

We might try to identify the split multiplets as bound states
with the light scalar.  This is clearly an appropriate description for scalar
mass $m>>\Lambda$.  Thus we might hypothesize that as we increase $m$, a
split multiplet evolves smoothly into a tower of bound states with  mass
splittings of $O(m)$ and decay rates $O(1)$.

It still may appear that there is a distinction between the picture
provided by spontaneously broken $U(1)^{N-1}\times S_N$ gauge theory,
and the more conventional confining description, which allows only
$U(1)^{N-1}\times S_N$ singlet states.
However, there is no invariant order parameter distinguishing
the two phases, the Higgs phase with $S_N$ spontaneously broken, versus the
confining phase.
This is true even in the theory containing fundamental matter, because the
center of $S_N$ is trivial.  In other words, the
eigenvalues of the adjoint scalar themselves transform under the $N$ (or
fundamental) of $S_N$ and every operator in the Higgs phase of the theory
will correspond to an $S_N$ singlet operator constructed by using the scalar.

Which description is physically more appropriate depends on the strength
of fluctuations in the $S_N$ sector.
In the language of spontaneously broken gauge theory,
a configuration contains domains
distinguished by different orderings of the $\vev{\phi_i}$,
and separated by domain walls.
The discrete gauge symmetry has the consequence that
the domain walls can end on additional string solutions, with energy
scale set by the symmetry breaking $a_i-a_j$.
The language of spontaneously broken gauge theory
is appropriate at energies low compared to this scale,
while at higher energies strings and domain walls can be created freely,
fluctuations in the $S_N$ are unsuppressed, and the confining language is
more appropriate.

The condition for weak coupling in the $S_N$ sector is the same condition,
$E<<m_W$, that we used earlier and motivated on general grounds of validity
of the effective Lagrangian.

Baryons will also exist in the theory and it is amusing to note that the
$U(1)^{N-1}$ charges of the quarks are such that the
flux tubes would form a chain running from the $i$'th quark to the $i+1$'st
quark in series.  This can be contrasted to other pictures proposed in the past
such as a flux tube with a `Y' junction.
Assuming the $W$ bosons exist as stable particles, such a state also exists but
is heavier for small $m$.

\newsec{The large $N$ limit.}

One of the original motivations for this work was to examine the
$N\rightarrow\infty$ limit of the theory, test the assumptions made in previous
work, and evaluate the hypothesis that large $N$ gauge theory is simpler than
finite $N$.  One question we need to address is the strength of the
coupling in the theory.   Simple probes of this are connected
expectation values of the gauge invariant field strength $\tr F^2$.
Standard large $N$ counting predicts $\vev{\tr F^2 ~ \tr F^2}_c ~ \leq O(1), ~
\vev{\tr F^2~ \tr F^2 ~\tr F^2}_c ~\leq O(1/N)$ and so on. If we calculate
these
expectation values at very long distance we can use the effective
Lagrangian \effL\  to evaluate them.  The field strength will be determined
by its abelian parts.

Let us first examine the naive semiclassical regime where the
electric couplings are weak and conventional perturbation theory is
valid.  To leading order the calculation is governed by the eigenvalues
of the kinetic term coefficient $\tau_{ij}$.  Call these eigenvalues
$\tau_i$.  It is not hard to see that in this regime $\tau_i \geq O(N)$.
The two point function is given by:
\eqn\twopt{\vev{\tr F^2 ~\tr F^2}_c \sim \sum_i {1\over {\tau_i^2}} \leq
O(1/N).}
This clearly satisfies large $N$ counting.  But this calculation also
illustrates a natural mechanism for its breakdown; namely, the
possibility of any
$\tau_i$  getting small.  But this is just what happens near points
where monopoles get massless, and so we expect large $N$ counting to
fail there.  Near these points the dual coupling is weak, so
perturbation theory for these quantities and hence \twopt\ is again
reliable.  This appears to be an example, of which several are already
known, of the lack of commutativity of the large $N$ and infrared limits
of a theory. It is the hope in ordinary confining gauge
theories that these limits will in fact commute.

We now can ask what is the domain of validity of the naive semiclassical
regime.  For concreteness we imagine moving on a line in the moduli
space that smoothly connects the extreme semiclassical regime with the
massless monopole point $\CC_0$. (Such paths are discussed in more
detail below.)  We can look for the breakdown of the semiclassical
regime by examining when the one loop approximation
becomes comparable to its first nontrivial correction, the one
instanton term which is of order $\Lambda^{2N}$.

Let us estimate the one instanton correction to $a_{Dm} =
\int_{\alpha_m} \lambda$.  The contour $\alpha_m$ surrounds the
cut connecting the branch points $x_{2m}$ and $x_{2m+1}$.
Semiclassically, i.e., for small $\Lambda$, each pair of branch points
surrounded by a $\gamma$ cycle are located very close to a zero of
$P(x)$, $x_{2i-1},x_{2i} \sim \phi_i$.  To the required accuracy we can
ignore the $x$ dependence  in every factor of $P$ except $(x-\phi_m)$ and
$(x-\phi_{m+1})$, replacing $x$ in these other factors by either
$\phi_m$ or $\phi_{m+1}$. The difference in results provides a measure
of the accuracy of the estimate.  Choosing $\phi_i$, we write $P(x) \sim
\half Q(x-\phi_i)(x-\phi_{i+1})$ where $Q=\prod_{j \neq i,i+1}
(\phi_i-\phi_j)$.  Using $\lambda = {1\over 2\pi i}{x\over y}{\p
P(x)\over\p x} dx$ we can write
\eqn\lamapprox{a_{Dm} \sim \int_{\alpha_m}
{{x(x-{\phi_m+\phi_{m+1}\over 2}) dx}\over
\sqrt{(\half(x-\phi_m)(x-\phi_{m+1}) )^2-\Lambda^{2N}/Q^2}}.}
This is closely related to the $SU(2)$ expression for $a_{Dm}$ with
the matching condition $\Lambda^4_{SU(2)}=\Lambda^{2N}/Q^2$.
The $SU(2)$ one instanton amplitude is $a_{Dm}^{1I}\sim
\Lambda^4/(\phi_{m+1}-\phi_m)^3$ So $a_{Dm}^{1I}$ is given by (assuming
$|\phi_{m+1}-\phi_m| >>\Lambda$)
\eqn\oneinstest{a_{Dm}^{1I} \sim {{\Lambda^{2N}}
\over{Q^2(\phi_{m+1}-\phi_m)^3}}.}

In the semiclassical region we have taken
$\phi_{ij}=v(i-(N+1)/2)\delta_{ij}$.  Inserting this in \oneinstest\
we find roughly
\eqn\est{\ a_{Dm}^{1I} \sim {{\Lambda^{2N}} \over{(N!)^2 v^{2N}}}
\sim ({{\Lambda} \over {Nv}})^{2N}.}
As $N \rightarrow \infty,~ a_{Dm}^{1I} \rightarrow 0$.  Note this is not
merely the $e^{-N}$ suppression of instantons. Such a suppression alone
(reflected in the factor $\Lambda^{2N}$) would allow this term to become
important for $v \sim \Lambda$.  The extra $N!$ makes this term
negligible until $v \sim 1/N$. (Of course this estimate is too
crude to detect possible power of  $N$ prefactors of $a_{Dm}^{1I}$.)
More generally, one instanton corrections
are negligible until the splittings between neighboring $\phi_i$ are
$\sim 1/N$.
 Physically, this suppression arises because of the $N$
different scales that are necessary to give all charged particles at
least $O(N^0)$ masses.  Instanton contributions can be understood as
associated with specific $SU(2)$ subgroups, and all of the mass scales
appear in the matching condition determining the coupling in a given
$SU(2)$ subgroup.  Thus the one loop expression for $a_{Dm}$ remains
exact except on a vanishingly small region of moduli space as $N
\rightarrow \infty$.

Of course this `vanishingly small' region includes $\CC_0$ and hence
is of vital interest for our study of confinement.  So we now turn to
the large $N$ physics in the immediate neighborhood of $\CC_0$.
A striking feature we have already alluded to is the emergence of
a very large hierarchy in scales near this point.
The BPS mass formula gives the $W_{ij}$ masses as
$m_{ij}=\sqrt{2}|a_i-a_j|$ with
$a_i={2N\over\pi}
(\sin {\pi i\over N}-\sin {\pi (i-1)\over N}) \rightarrow
2\cos {\pi i\over N}$ at large $N$.
The smallest
mass is for $W_{i,i+1}$ which for large $N$
goes as $m_{i,i+1}\rightarrow {2\pi\sqrt{2}\over N}\sin {\pi i\over N}$.
This is of order $\sim 1/N$ for $i$ near the center of the band,
and $\sim 1/N^2$ for $i$ near the edge of the band.

The mass formula does not guarantee the existence of these $W$ states.
There are hyperplanes in moduli space on which particular BPS saturated
states become unstable to decay, and thus the states need not be present
in the spectrum on both sides \refs{\vafaetal,\sw}.  When the ratio of the
central charges
$Z(q,h)=a_i q^i+a_{Di} h^i$ associated with two particles becomes real,
$Z(q',h')/Z(q,h)\in\BR$, the particle with charges $(q+q',h+h')$ is
neutrally stable to decay into particles $(q,h)$ and $(q',h')$.
We should check that these light $W$s are
are present in the spectrum near $\CC_0$.
In fact they cannot be present at every point near $\CC_0$.
The light $W$'s are nonlocal
with respect to the massless monopoles $M_m$ and their mass formulae
are multivalued around the massless monopole point, as in \sw.\footnote*{
We thank N. Seiberg for pointing this out to us.}
Thus for each light $W$ there must be a hyperplane passing through $\CC_0$ on
which it goes unstable.

This does not exclude the possibility that there exists a path connecting
$\CC_0$ with the semiclassical regime on which the $W$'s remain stable.
We have found such a path, generalizing a similar path in \sw\ for $SU(2)$.
It simply interpolates between the two regimes as follows:
\eqn\evpath{\phi_i(t) = (1-t)\phi_i|_{\CC_0} + t({N+1\over 2}-i).}
For $N>5$ and $t<<1$ the double zeroes of $P^2-1$ are split slightly and remain
on the
real axis.
If we can show that for every $t<1$ the zeroes of $P^2-1$ remain distinct, they
must remain on the real axis, and the assignment of branch cuts and periods can
be held fixed.  We have shown this numerically for $N\le 21$, and the behavior
of the zeroes is sufficiently simple that we are convinced that this persists
for arbitrary $N$.  The accompanying plot shows the motion of the zeroes
of $P^2-1$ as a function of $t$ for $N=7$.

\medskip

\ebox{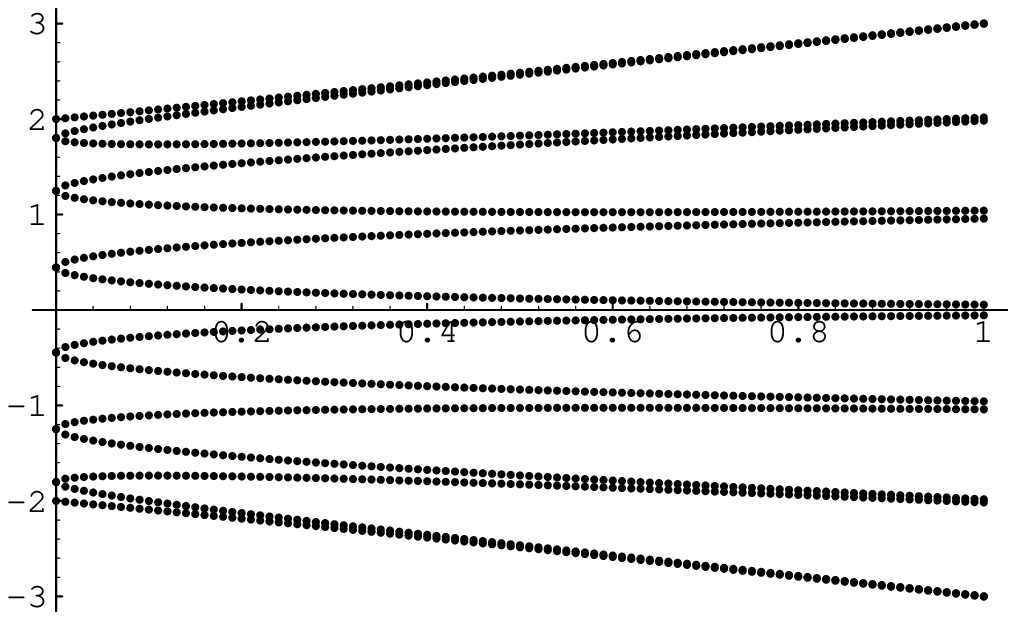}{}

\medskip

Thus the $a_{Dm}$ will remain imaginary and the $a_m$ will remain real for all
$t$, it will be impossible for electrically charged states to decay into
magnetically charged states for any $t$, and the presence of the $W$'s in the
semiclassical regime implies the existence of light electrically charged states
near $\CC_0$.  We have not yet proven the existence of a
particular charged state.  Because all $a_m$ are real along the
trajectory, all $W$'s are potentially neutrally stable.
Nevertheless the lightest $W$'s will exist all along the trajectory.
This follows from the fact that the ordering $a_m>a_n$ for all $m>n$ is
preserved along the trajectory, for which the argument is simply that the $a_m$
are real, so that if the ordering changes there must be a point in moduli space
at which $a_m=a_{m+1}$, but this would imply the existence of a massless $W$
and a singularity on moduli space not associated with a degeneration of the
curve, which is impossible.
Given this ordering, the particles $W_{m,m+1}$ are necessarily stable all along
the trajectory.

Although the $W$ bosons clearly exist in the $\CN=2$ theory,
the question for the $\CN=1$ theory is subtle, because we are working so close
to a surface on which they go unstable.  We cannot prove it with our present
techniques, and will assume it is true.

Our effective Lagrangian is only valid for momenta small compared to the
lightest $W$ mass, $p \leq 1/N^2$.  This is another indication of the
noncommutativity of the large $N$ and infrared limits in this theory.
There should be a signature of this scale in the quantities we have
already calculated, since the logarithmic running of the dual magnetic
coupling at the edge of the band due to monopole loops should only start
at scales below this mass.  In fact we see in \perstufffourb\ precisely
this effect.

There are a number of phenomena occurring near $\CC_0$.  First, as we
have seen, instanton effects can become important roughly when eigenvalue
splittings $\Delta\phi \sim \Lambda/N$. This indicates the initial
breakdown of the semiclassical regime.  Also, near $\CC_0$ the monopoles
become light, at some point becoming comparable in mass to the
lightest $W$ they couple to. {}From \perstufftwo\ and the inverse of
\Bdiag\ we see that to give the monopoles such masses $a_{Dm}\sim
(\Lambda/N)\sin \pi m/N$, we need to vary $\CC_0$ as $\delta\phi_i\sim
(\Lambda/N)\cos \pi (i-\half)/N$.  This is a variation of $\phi_i$ which
is $O(1/N)$ compared to its value at $\CC_0$.  We see that the
masslessness of the monopoles is due to delicate cancellations, a
mathematical explanation of the appearance of multiple scales.
In this region one can study the combined dynamics of
electrically and magnetically charged particles as both their masses are
made arbitrarily light with their ratio fixed.  The first bit of this
physics is the varying logarithmic cutoff of \perstufffourb.
A more detailed discussion will follow in the next section.

We now turn our attention to the confining phase by applying the $\CN=2$
breaking perturbation $Nm Tr A^2$ discussed earlier. The light $W$s
dramatically affect the domain of $m$ we can control. They cause flux
tube breaking and coupling of the different $U(1)$ sectors with no
suppression at energy scales above their mass, making the theory
unmanageable.  So we are restricted to regime where the confining scale
(e.g., a glueball mass) is lighter than the $W$ scale. This requires
taking $m \sim \Lambda/N^4$.  Naively one might have expected to be able
to take $m \sim \Lambda$.  We see that as $N \rightarrow \infty$ the
abelian flux tube picture of confinement is valid on a vanishingly small
part of the $m$ axis.

In this region there is no reason to think that three Wilson loop
expectation values are other than order one, and the glueball couplings
described by the effective Lagrangian are manifestly so.  If
conventional large $N$ physics holds in the pure $\CN=1$ SYM theory
obtained as $m \rightarrow \infty$ a complicated but smooth rearrangement
of the theory must take place. There certainly will be complicated
dynamics occurring for $m \geq \Lambda/N^4$. Whether conventional
expectations are realized we cannot say from this analysis. It is
conceivable that order one couplings
could persist as $m \rightarrow \infty$.

If conventional expectations are valid, they would probably hold
until string tensions become of order  light $W$  masses.  Since these masses
go to zero as $N \rightarrow \infty$ the region of validity of
conventional expectations would occupy the whole $m$ axis at $N=\infty$.
There would be a sharp transition between the coulomb $\CN=2$ theory
described by one loop physics and the confining theory described
by large $N$ lore.  The light abelian monopole region that smoothly
interpolates between them  would have disappeared.

\newsec{A scaling trajectory.}

So far, our discussion of the large $N$ limit has been largely qualitative.
We saw that as we move in the $\CN=2$ moduli space towards $\CC_0$, many things
happen.  We start with a simple $U(1)^{N-1}$ gauge theory in which many
calculations are exact at one loop.  Gradually instantons turn on, and at the
same time monopoles become light, eventually crossing the $W$ masses to become
the lightest degrees of freedom.  They will then drive the beta function and
cause
the $U(1)$ gauge factors to decouple in the monopole basis, different from the
original basis in which $\vev{\phi}$ was diagonal.
Our analysis suggested that all of this happened in the scaling regime
$|\phi_i-\phi_{i+1}|\sim 1/N$.

We will now study the relationship between these phenomena more
systematically by introducing the following scaling trajectory
for the eigenvalues of $\vev{\phi}$:
\eqn\scaletraj{\phi_i(s) = e^{s/N}~\phi_i|_{\CC_0} =
e^{s/N}~2 \cos {\pi(i-\half)\over N}~ .}
Introducing the scaled coordinate $z=e^{-s/N}x$ we find the polynomial
is of the form $P(x,s)=e^s P(z)$ and the curve is
given by $y^2=e^{2s}(P^2(z)-e^{-2s})$ where $P(z) \equiv P(z)|_{\CC_0}$.
As for our previous trajectory, the branch point evolution is smooth:
taking $z=2\cos\theta$ we have $P(z)=\cos N\theta$ and the branch points
are at $\cos N\theta=\pm e^{-s}$.

The trajectory is entirely in the scaling regime and even with $s$ large
the outlying $|\phi_i-\phi_{i+1}|$ remain $O(1/N^2)$.
However its essential property is that for $s$ large, the length of the
$\gamma$ cycles shrinks to zero (in terms of $x$, as $2e^{-s}$) and thus we
recover semiclassical results.

The form $\lambda$ becomes
\eqn\lambdascale{\eqalign{
\lambda &=  {{e^{s/N}}\over{2 \pi i}}
{{z   d z} \over {\sqrt{P(z)^2-
e^{-2s}}}}{{\partial P}\over{\partial z}}\cr
&={{1}\over{\pi i}}
{{e^{s/N}\cos \theta~ d(\cos N \theta)}\over{
\sqrt{\cos^2 N \theta - e^{-2s}}}}.}}
The large $N$ simplification occurs because the integrals
over the $\alpha$ and $\gamma$ cycles only cover a range of
$\theta \sim 1/N$.
Thus we can expand the $\cos \theta$ factor to isolate the leading $N$
dependence.

\fig{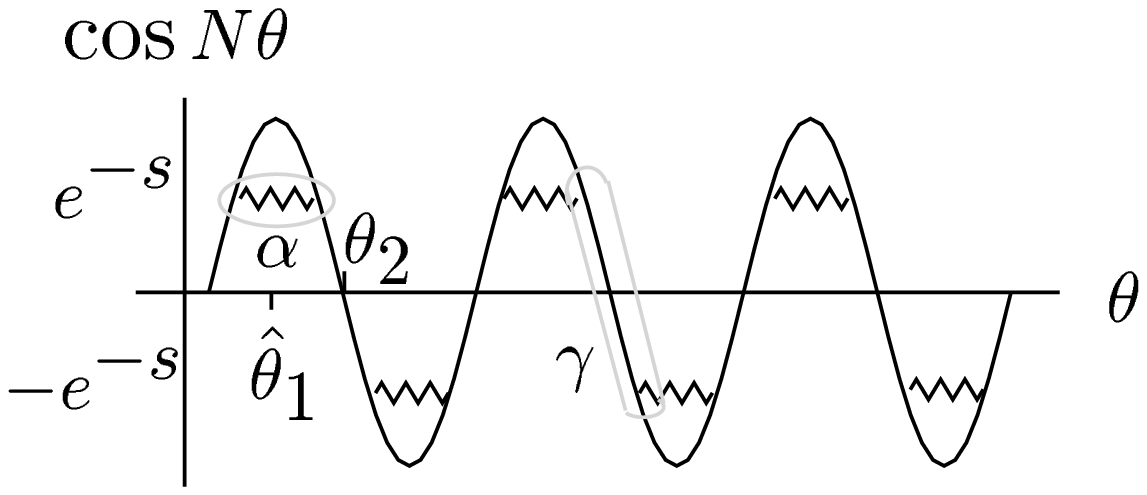}{}

For $a_i$ and $a_{Dm}$ we obtain%
\footnote*{The integral $I_1(s)$ is most easily evaluated by
using partial integration to establish $I_1(s)=-I_1(-s)$. The
only solution to this compatible with the analytic properties of $I_1$
is $I_1=s$.  We are grateful to Alyosha and Sasha Zamolodchikov for
providing this argument.}
(using $u=\cos N \theta$)
\eqn\scaleresulta{\eqalign{ a_i
&= {{{e^{s/N}~ 2 \cos \theta_i}\over{\pi}}} \int_{-e^{-s}}^{e^{-s}}
{{du}\over{\sqrt{e^{-2s}-u^2}}} + O({1\over N^2})~. \cr
&= {e^{s/N}~ 2 \cos \theta_i}}}
and
\eqn\scaleresultad{\eqalign{a_{Dm}
&= {{4 i  \sin \hat \theta_m}\over{ \pi }}e^{s/N} ({{1}\over{N}} I_1(s)
 + O({1\over N^3})) \cr
&= {4 i s e^{s/N}\over N}\sin \hat \theta_m ~. \cr
I_1(s)&= \int_{u =e^{-s}}^{u =1} {{d u~ \arccos u }\over {\sqrt{u^2 -
e^{-2s}}}} \cr
&= s~.}
}

We see that we can make monopoles much heavier than $W$'s
along this trajectory.  Monopole masses are comparable
to the lightest $W$ they couple to for $s \sim 1$.
One loop
contributions going like $\log (a/\Lambda)$ will show $s$ dependence $\sim
s/N$, and we see from \scaleresultad\ that at leading order in $N$
the monopole and $W$ masses are given {\it exactly} by the one loop result,
all the way to the massless monopole point!
(We will see below that this is a special property of this trajectory and
more generally trajectories $\delta\phi_i\sim \cos \pi k(i-\half)/N$
with $k\sim O(N^0)$.)

To compute $\tau$, we might use
\eqn\partlambda{{\p\lambda\over\p\phi_i} =
{e^{s/N}\over 2\pi}{\sin\theta\over \cos\theta-\cos\theta_i}
{\cos N\theta\over\sqrt{\cos^2 N\theta-e^{-2s}}}d\theta.}
There is a subtlety in the large $N$ expansion of this form,
due to the pole in the prefactor at $\theta=\theta_i$.
Although the pole itself is cancelled by a zero of $\cos N\theta$,
it is not justified to treat the prefactor as slowly varying.

We will deal with this by using the basis of variations
\eqn\newint{{\p\lambda\over \p t_k} \equiv
\sum_{i=1}^N {\p\lambda\over\p\phi_i} \cos {\pi k(i-\half)\over N},}
We will find that using this basis for `electric' quantities and the basis
$\sin k\hat\theta_m$ for `magnetic' quantities,
the period matrices become simple and the kinetic term diagonal, for all $s$
and at finite $N$.

Before doing this in detail, let us suggest why this gives
simple results.  From \Bdiag\ and related formulas, it is clearly an
appropriate basis for the massless monopole point $s=0$.
On the other hand, the large $s$ limit
is expected to reproduce the semiclassical
results $\p a_i/\p\phi_j=\delta_{ij}$ and
$\p a_{Di}/\p\phi_j=\p^2 \CF/\p a_i\p a_j$.
{}From \Foneloop\ and \scaleresulta\ this is
\eqn\Fonelscale{\eqalign{
{\p a_{Di}\over\p\phi_j}\bigg|_{\rm one\ loop} =
\tau^{(1)}_{ij} &=
{i\over 2\pi}\left[
\delta_{ij} \sum_{k\ne i} t_{ik} - (1-\delta_{ij}) t_{ij} \right]\cr
t_{ij} &= {2s\over N}+\log(2\cos\theta_i-2\cos\theta_j)^2.}}
The point is that the one-loop kinetic term is also diagonal
in the cosine basis, by results similar to those in the appendix,
making it possible for $\tau$ along the entire trajectory to be diagonal.
Let us estimate its eigenvalues in the large $N$ limit:
the difference $t_{i+1,j}-t_{i,j}\sim B_{m=i,j}+O(1/N)$ and so
\Bdiag\ implies that
$t_{ij}$ will have eigenvalues
$t_\kappa \sim {1\over 2\sin {\pi k/N}}$.
(There are corrections to this estimate, of $O(k/N)$).

We proceed to compute the periods of
\eqn\newint{\eqalign{ {\p\lambda\over \p t_k} &\equiv
\sum_{i=1}^N {\p\lambda\over\p\phi_i} \cos {\pi k(i-\half)\over N}\cr
&=
{1 \over 2\pi}
{N \sin (N-k)\theta\over\sqrt{\cos^2 N\theta-e^{-2s}}}d\theta.}}
We have ($a = \arccos e^{-s},~ b=\arcsin e^{-s},~ \kappa=k/N$)
\eqn\newint{\eqalign{ A_{jk}(s)&={\p a_j\over \p t_k} =
{(-1)^{j-1}\over \pi} \int_{-b}^{b}d\theta~
{\sin ((N-k)(\theta_j+\theta/N))
\over\sqrt{e^{-2s}-\sin^2 \theta}}\cr
&= {1 \over \pi} \int_{-b}^{b}d\theta~
{\cos k\theta_j\cos (1-\kappa)\theta + \sin k\theta_j\sin(1-\kappa)\theta
\over\sqrt{e^{-2s}-\sin^2 \theta}}\cr
&= F(\kappa,s)~\cos k\theta_j \cr
F(\kappa,s)&= {1 \over \pi} \int_{-b}^{b} d\theta {{\cos
(1-\kappa)\theta} \over {\sqrt{e^{-2s}-\sin^2 \theta}}}\cr
&\rightarrow {1\over\pi}\sin{\pi\kappa\over 2} (-\log s)
+ 1 - O(\kappa) \qquad\qquad {s\rightarrow 0} \cr
&\rightarrow 1 +{\kappa(2-\kappa)\over 4}e^{-2s}+\ldots \qquad\qquad\qquad
{s\rightarrow\infty} }}
and similarly
\eqn\newintd{\eqalign{B_{mk}(s)={\p a_{Dm}\over \p t_k} &=
G(\kappa,s)~\sin k\hat\theta_m \cr
G(\kappa,s) &=
{1 \over \pi} \int_{-a}^{a}d\theta'~
{\cos(1-\kappa)\theta'
\over\sqrt{\cos^2 \theta'-e^{-2s}}} \cr
&\rightarrow 1 +{\kappa(2-\kappa)\over 2}s+\ldots \qquad\qquad {s\rightarrow 0}
\cr
&\rightarrow 2s \sin {\pi\kappa\over 2} + 1 - O(\kappa)
+\ldots\qquad\qquad
{s\rightarrow\infty} .}}

We can now compute $\tau_{Dmn}(s)$. Using \qsin\ we see that
 $\tau_{Dmn}(s)$ is diagonal in the $\sin k\hat\theta_m $ basis:
\eqn\diagtaud{\eqalign{\tau_{Dmn}(s) = \sum_k A_{mk}B^{-1}_{kn} &=
\sum_{i,k} q^{-1}_{mi}
A_{ik}B^{-1}_{kn}, \cr
\sum_{n}\tau_{Dmn}(s)\sin k\hat\theta_n &=
\tau_{D\kappa}(s)\sin k\hat\theta_m}}
with eigenvalues
\eqn\ev{\tau_{D\kappa}(s) = {1\over 2\sin{\pi\kappa\over 2}}
{F(\kappa,s)\over G(\kappa,s)}.}
We verify that $\tau_{D\kappa}(s) \rightarrow
-{1\over{2\pi}}\log s$, independent of $\kappa$, as $s \rightarrow 0$.
Similarly
\eqn\diagtau{\eqalign{\tau_{ij}(s) &= \sum_{m,k} h^{-1}_{im}B_{mk}A^{-1}_{kj}
 \cr
&\sum_{j}\tau_{ij}(s)\cos k\theta_j =
\tau_{\kappa}(s)\cos k\theta_i\cr
\tau_{\kappa}(s) &= {1\over 2\sin{\pi\kappa\over 2}}
{G(\kappa,s)\over F(\kappa,s)}.}}
These formulae give an exact description
of the physics along the scaling trajectory from the semiclassical
domain all the way to $\CC_0$, even at finite $N$.

Note the crucial fact that the eigenvalues $\tau_{\kappa}(s)$ are {\it not}
simply the inverses of the eigenvalues $\tau_{D\kappa}(s)$!
The change of basis from electric ($ij$) to magnetic ($mn$) couplings
introduces an extra $1\over 4\sin^2{\pi\kappa\over 2}$.
This will have important consequences and seems quite mysterious;
it may be relevant that (for example) in the
magnetic basis, the $W$ bosons responsible for the one loop contribution to
$\tau$ have charges with alternating signs (such as $(1\ -1\ -1\ 1)$~), and
in summing their loop effects there will be large cancellations.

Let us consider the leading correction to the logarithm from the monopole beta
function at small $s$, which come from the constant terms in $F$ and $G$:
\eqn\leadcor{\tau_{D\kappa}(s) =
-{1\over{2\pi}}\log s + {1\over 2\sin{\pi\kappa\over 2}} + O(1)+
\ldots .} For $k\sim O(N^0) ~ (\kappa \sim 1/N)$ and $s\sim O(N^0)$, this
is a large correction, of $O(N)$.  It is a threshold effect and
eventually the monopole-induced running of the coupling will overwhelm
it, but we need to go to scales $s\sim e^{-N}$ for this to happen.  Thus
these effects are quite important at large $N$.

The same constants are responsible for the $s$-independent term at large
$s$ in
\eqn\leadcort{\tau_{\kappa}(s) =
2s + {1\over 2\sin{\pi\kappa\over 2}} + O(1) + \ldots .}
We see by comparison with \Fonelscale\ that this is just
$\tau_{\rm one\ loop}$.
Rather magically, the low $k$ `modes' manage to be simultaneously weak
in electric and magnetic variables.

Since the constant term in $\tau_{D\kappa}$ has the same form, the form
of $\tau_D$ in the monopole basis can be inferred by analogy to $\tau$:
\eqn\tauDfin{\tau_{Dmn} \sim -{i\over{2\pi}}\delta_{mn} \log s +
{i\over{2\pi}} \log (2\sin\hat\theta_m-2\sin\hat\theta_n)^2
+ O(1) + \ldots .}
This matrix is not diagonal and so a `magnetic observer' would see
peculiar transitions between the $m$ quanta.  This observer might
posit the existence of light solitons carrying dual charge in four $m$
factors (as mentioned above) that would mediate these transitions.

The similarity of \tauDfin\ to the one-loop electric $\tau$
is very intriguing and leads us to wonder if there is any sense in which
it is a one-loop effect.  The theory is certainly much more
`self-dual' than one would have guessed.

\fig{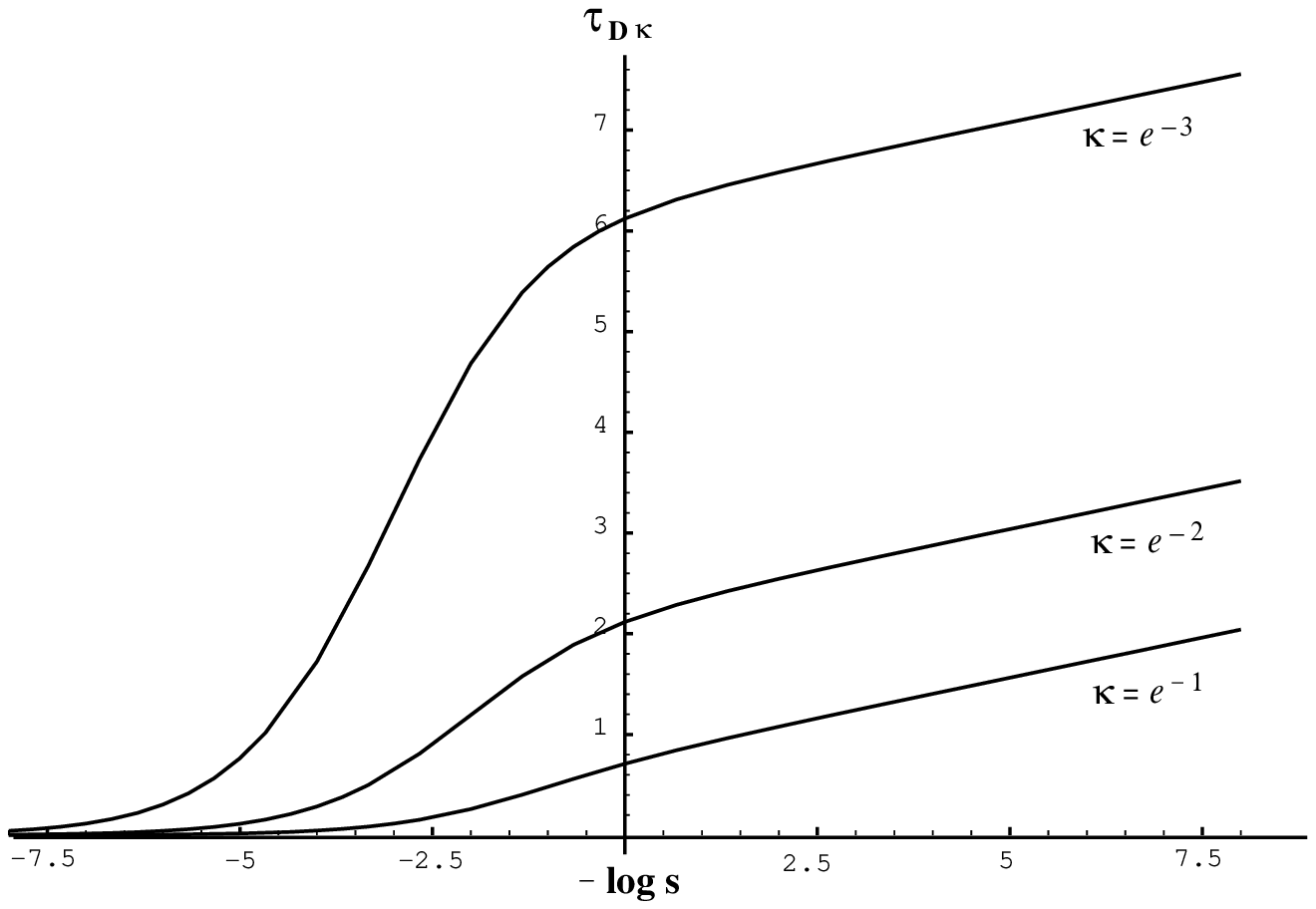}{}
We now plot the scaling functions to illustrate their structure.
Note that the
$- {1\over{2\pi}}\log s$  behavior characteristic of the monopole
one loop beta function sets in around $s=1$ uniformly in $\kappa$.
This is just at the energy scale set by the inverse
`size' of the monopole, as discussed earlier.
The  large residual $s$ independent term in \leadcor\ is responsible
for the $\kappa$ dependent offsets of the curves.
It is clear from the graph that one must go to $s \sim e^{-1/\kappa}$
before this offset is negligible.

The off diagonal terms in \tauDfin\ influence the energies of states
in the confining phase obtained by adding the $\CN=2$ breaking
perturbation $Nm Tr A^2$.  The basic structure of the flux tubes
is determined by the monopole condensate which is determined
{}from the superpotential alone and hence does not depend on
$\tau_{Dmn}$.  It is still given by \nonesol.  The string tensions
do depend on $\tau_{Dmn}$ and will be affected by the mixings
described by \tauDfin.  We have not yet studied this problem,
but we have examined the closely related problem of the
dual photon masses, by numerically diagonalizing the mass matrix.
We found that the
quantitative relation \massfor\ is altered but that the scales of the
masses remain unchanged.  We expect the same for the string tensions.
Note that these changes are significant for $m \sim e^{-N}$, far
below the energies where string breaking due to light $W$ pairs
occur ($m \sim 1/N^4$).

We remind the reader that some quantities ($a_i$ and $a_{Dm}$)
were exact at one loop along the scaling trajectory.
This occurred because the corrections to the one-loop results for their
derivatives, from \newint-\newintd, were suppressed by powers of $\kappa$,
so would happen for any trajectory with $\kappa\sim 1/N$.
If the same thing happened in other large $N$ (supersymmetric) theories,
for even a single
trajectory into the true ground state (here the massless monopole point),
it would have the important consequence that the ground state could be found
just knowing the one-loop effective action.

\newsec{Conclusions}

Using the solution of $\CN=2$ supersymmetric $SU(N)$ gauge theory
of \refs{\sw,\AF,\KLTY}, we have given an overview of some of the
physics visible in the low energy effective Lagrangian.
In particular we studied the $\CN=1$ theory obtained by giving a mass
to the chiral superfield, and extended Seiberg and Witten's explicit
derivation of a
monopole condensation model of confinement for $SU(2)$ to $SU(N)$.

The Lagrangian relevant for this confining theory
is a weakly coupled
$U(1)^{N-1}$ gauge theory with one light monopole hypermultiplet
in each factor. The $S_N$ discrete gauge symmetry that permutes
the $U(1)$ factors is {\it spontaneously broken}.  This phenomenon
pervades the physics of this system.  In particular, there is a spectrum
of different string tensions in the different factors, $\kappa_n \sim
m \Lambda N\sin {\pi n\over N}$ and a spectrum of different light $W$
boson masses $\sim {{\Lambda}\over {N}} \sin {\pi n\over N}$ that
connect the factors.  This differs in many ways from our expectations
about the pure $\CN=1$ theory.  It is  possible for a dramatic
but smooth rearrangement to take place for $m \sim \Lambda$ that will
connect these two descriptions.  Perhaps this rearrangement can be
understood as a smooth crossover from a regime of spontaneously
broken $S_{N}$ gauge symmetry to one where the $S_N$ gauge symmetry
is `confined.'

We studied the large $N$ limit in detail.  The hierarchy of scales
present for finite $N$ becomes very large, as the lightest $W$ mass
$\sim \Lambda/N^2$.  We found one signature of these very light
particles in the monopole `size' which controls the energy scale of
onset of perturbative monopole coupling constant renormalization.
Another very surprising signature is the off-diagonal terms in
\tauDfin\ which cause a coupling of the different magnetic factors.
Understanding this phenomenon from the magnetic point of view,
presumably as the result of light electric solitons in a weakly
coupled magnetic theory, would be very interesting.

At large $N$ the one loop result becomes exact almost everywhere
on the $\CN=2$ moduli space.  We identified a scaling regime very close
$(\sim 1/N)$ to the massless monopole point $\CC_0$ where instanton and
monopole effects survive the large $N$ limit and were able to give simple
exact formulas for monopole and $W$ masses and coupling constants
along a scaling trajectory through this regime.

The coupling almost everywhere on the $\CN=2$ moduli space is $1/N$,
as expected, but near $\CC_0$ becomes large, contradicting standard
large $N$ lore.  The infrared divergence of the effective $U(1)$
couplings, which does not commute with the large $N$ limit, is the source
of this.  Such a phenomenon would seem to apply to any continuous monopole
condensation picture of confinement, supersymmetric or otherwise.
The region of this violation goes to zero as $N \rightarrow \infty$.

In the confining `almost $\CN=2$' theory, the abelian flux tube picture
breaks  down because of $W$ pair creation when confining scales become
comparable to $W$ scales.
This occurs for $m \sim \Lambda/N^4$, i.e., a vanishingly small region
of the $m$ axis.  Conventional large $N$ lore could be recovered in
a smooth, complicated crossover we cannot control.  As $N \rightarrow
\infty$ the region described by nontrivial light monopole physics
shrinks to zero, possibly leaving an abrupt transition between one loop
physics and large $N$ pure $\CN=1$ SYM behavior.

Some features of this model are reminiscent of standard
hadronic phenomenology while others
look very different (such as the weakly coupled glueballs), but
overall we find the results encouraging for the idea of using
supersymmetric gauge theory as a solvable starting point for attempts to
model QCD dynamics.
Surely interesting supersymmetric analogs of many other problems
of strongly coupled gauge theory and QCD physics can be found,
and we believe it will be very fruitful to study them using these techniques.

\medskip
It is a pleasure to thank Philip Argyres, Tom Banks, Dan Friedan,
Ken Intriligator,
Wolfgang Lerche, John Preskill, Lisa Randall, Nati Seiberg and
Alyosha and Sasha Zamolodchikov for
valuable conversations.

We are informed by Greg Moore that he and Mans Henningson have independently
obtained some of these results.

This research was supported in part by DOE grant DE-FG05-90ER40559, NSF
PHY-9157016 and the A. P. Sloan Foundation.

\appendix{A}{Trigonometric identities.}

In section 2 we use
\eqn\Bdiag{\sum_{m=1}^{N-1} B_{mi} \sin {\pi km\over N} =
 \cos {\pi k(i-\half)\over N}}
which can be checked by writing
($z_m=e^{i\hat\theta_m}$, $w_i=e^{i\theta_i}$)
\eqn\diagtwo{B_{mi} = {1\over iN}\left(
{1\over 1-w_iz_m}-{1\over 1-w_iz_m^{-1}}\right).}
We convert the sum to run from
$0$ to $2N-1$, for which $\sum_m e^{\pi i k m/N}=2N\delta_{k,0({\rm mod}2N)}$,
by taking a sum $\sum_{m=1}^{N-1}$ involving the second
term in \diagtwo\ and taking $m\rightarrow 2N-m$.  Then
\eqn\diagthree{\eqalign{\sum_{m=1}^{N-1}B_{mi} \sin {\pi km\over N} &=
{1\over 2N}
\sum_{m=1}^{2N-1}
{w_i^{-1}\over w_i^{-1}-z_m}\left(e^{-i\pi km/N}-e^{i\pi km/N}\right)\cr
&={w_i^{k}\over 1-w_i^{2N}} - {w_i^{2N-k}\over 1-w_i^{2N}}\cr
&={w_i^k+w_i^{-k}\over 2}.}}

Using \Bdiag, the orthogonality of the basis $\sin \pi k m/N$
under $\sum_{m=1}^{N-1}$, and the orthogonality of the basis
$\cos {\pi k(i-\half)/N}$ under $\sum_{i=1}^{N}$, we also have
\eqn\Bdiagtwo{\sum_{i=1}^{N} B_{mi} \cos k\theta_i =
 \sin k\hat\theta_m.}

In section 5 we use the change of basis
\eqn\qsin{\sum_m q^m_i \sin k \hat\theta_m =
\sin k \hat\theta_{m=i}-\sin k \hat\theta_{m=i-1} =
2\sin {k \pi\over {2N}} \cos k\theta_i}

We next evaluate the sum
\eqn\fsum{f_k(\theta) = {1\over N}\sum_{j=1}^N
{\sin\theta\over \cos\theta_j-\cos\theta} \cos k\theta_j =
{\sin (N-k)\theta\over \cos N\theta}.}
First, for $k=0$, the sum is
\eqn\fsumtwo{{1\over N}{\p\over\p\theta}\log\prod_j
(\cos\theta_j-\cos\theta) =
{1\over N}{\p\over\p\theta}\log P(x) = \tan N\theta.}
For $k$ integer, the sum is a periodic function of $\theta$
with residues $\cos k\theta$ at the points $\theta=\theta_j$,
so it must take the form
\eqn\fsumthree{f_k(\theta) = \tan N\theta \cos k\theta + {\rm finite}.}
{}From the above,
the finite part is known at the points $\theta=\hat\theta_m$
to be $-\sin k\theta$.  This is true of the function
\eqn\fsumfour{f_k(\theta) = \tan N\theta \cos k\theta - \sin k\theta}
(equal to \fsum) plus any periodic function vanishing
at all of the $\hat\theta_m$, such as $\sin 2Nl\theta$.
Such terms can be excluded by considering the derivative $d/d\theta$
at $\theta=0$.

\bigskip

\listrefs
\end